\documentclass{aastex701}

\usepackage{rotating}
\usepackage{amsmath}
\usepackage{soul}
\usepackage{lipsum}
\usepackage{amsthm,amsmath,amssymb}

\hypersetup{linkcolor=red,citecolor=blue,filecolor=cyan,urlcolor=magenta}

\begin{document}

\title{BSN-V: The First Detailed Light Curve Modeling of Eight Totally Eclipsing Contact Binary Stars Using Ground-Based and TESS Observations}

\author[0000-0002-0196-9732]{Atila Poro}
\altaffiliation{atila.poro@obspm.fr}
\affiliation{LUX, Observatoire de Paris, CNRS, PSL, 61 Avenue de l'Observatoire, 75014 Paris, France}
\affiliation{Astronomy Department of the Raderon AI Lab., BC., Burnaby, Canada}
\email{atila.poro@obspm.fr}

\author[0000-0003-1263-808X]{Raul Michel}
\altaffiliation{rmm@astro.unam.mx}
\affiliation{Instituto de Astronom\'ia, UNAM. A.P. 106, 22800 Ensenada, BC, M\'exico}
\email{rmm@astro.unam.mx}

\author[0000-0002-9761-9509]{Francisco Javier Tamayo}
\affiliation{Facultad de Ciencias F\'{\i}sico-Matem\'aticas, UANL, 66451 San Nicol\'as de los Garza, NL, M\'exico}
\email{francisco.tamayomy@uanl.edu.mx}

\author[0000-0002-0192-215X]{Mahya Hedayatjoo}
\affiliation{Department of Physics, Iran University of Science and Technology, 16846 Tehran, Iran}
\email{mhedayatjoo1383@gmail.com}

\author[0000-0002-7348-8815]{Hector Aceves}
\affiliation{Instituto de Astronom\'ia, UNAM. A.P. 106, 22800 Ensenada, BC, M\'exico}
\email{aceves@astro.unam.mx}

\author[0000-0002-1972-8400]{Fahri Alicavus}
\affiliation{Çanakkale Onsekiz Mart University, Faculty of Arts and Sciences, Department of Physics, 17020, Çanakkale, Türkiye}
\affiliation{Çanakkale Onsekiz Mart University, Astrophysics Research Center and Ulupnar Observatory, 17020, Çanakkale,
Türkiye}
\email{fahrilcvs@gmail.com}

\begin{abstract}
This study broadens our comprehensive investigation of total-eclipse W Ursae Majoris-type contact binaries by analyzing eight additional systems, continuing our previous research. Multiband $BVR_cI_c$ photometric data were obtained at an observatory in Mexico, from which new times of minima were determined. All target systems also had available space-based TESS time-series data. Orbital period variations were studied for eight target systems, showing either linear or parabolic trends. The target systems exhibiting parabolic trends demonstrated a sustained decrease in their orbital periods over time. We modeled the light curves utilizing the PHOEBE Python code in combination with the BSN application. We revisited the relationship between orbital period and the temperature of the hotter component in contact binary systems using an empirical approach. Our analysis identified a clear break at $P=0.27$ days, separating the systems into two distinct groups for orbital periods shorter than 0.6 days. Following the determination of stellar extinction, absolute parameters for seven systems were estimated employing parallax measurements from Gaia DR3. Based on the components' effective temperatures and masses, the systems were classified into A- and W-subtypes. Their evolutionary states were illustrated using mass–radius and mass–luminosity diagrams.
\end{abstract}


\section{Introduction}
\label{sec1}
Contact binaries of the W Ursae Majoris (W UMa) type feature short orbital periods in which both stellar components fill their Roche lobes and share a common convective envelope (\citealt{kopal1959close}, \citealt{lucy1968structure}, \citealt{lucy1968light}, \citealt{lucy1979observational}, \citealt{mochnacki1981contact}). These systems are mostly composed of late-type, low-mass stars, typically of spectral types F, G, K, and occasionally M. The shared envelope promotes highly efficient energy and mass exchange, often resulting in similar surface temperatures and nearly equal eclipse depths in the observed light curves (\citealt{kuiper1941interpretation}, \citealt{2005ApJ...629.1055Y}).

Despite extensive studies over the past eight decades (\citealt{kuiper1941interpretation}, \citealt{guinan1988formation}, \citealt{bradstreet1994asp}, \citealt{2017RAA....17...87Q}), key questions about the structure and long-term development of these binaries remain unresolved. One widely accepted evolutionary path suggests that W UMa-type binaries form from detached short-period binaries through angular momentum loss, particularly due to magnetic braking (\citealt{1982AA...109...17V}, \citealt{2006AcA....56..347S}, \citealt{2017RAA....17...87Q}). Over time, these systems may evolve into rapidly spinning single stars, such as blue stragglers or FK Com-type stars, following a merger event, although empirical evidence for such outcomes remains limited (\citealt{1995PASP..107..648R}, \citealt{2011AA...528A.114T}).

The orbital period distribution for contact binaries exhibits a strong peak around 0.27 days, with a pronounced short-period cutoff near 0.2 days. While this lower limit has been commonly attributed to constraints imposed by angular momentum loss or structural instability in low-mass components (\citealt{2012MNRAS.421.2769J}), its exact value and physical origin are still being actively investigated. Moreover, observed period changes can offer valuable insights into internal processes such as mass transfer and angular momentum evolution (\citealt{li2004structure}, \citealt{2006AJ....132.2260H}, \citealt{2016AJ....152..120H}, \citealt{2013ApJS..209...13Q}).

Photometric asymmetries in contact binaries, most notably the O’Connell effect, where the two maxima in the light curve differ in brightness, are another important diagnostic. These asymmetries are typically attributed to magnetic spots, although additional mechanisms like mass transfer-induced hot spots or surrounding material may also play a role (\citealt{1951PRCO....2...85O}, \citealt{milone1968peculiar}, \citealt{davidge1984study}). Accurately interpreting these effects is essential, as surface inhomogeneities can distort photometric solutions.

Reliable determination of physical parameters, such as mass ratio, fill-out factor, and temperature ratio, is fundamental to our understanding of contact binary evolution. While spectroscopic radial velocity data are often unavailable due to observational challenges, studies have shown that for systems exhibiting total eclipses, mass ratios inferred from photometry can be consistent with spectroscopic values (\citealt{1972MNRAS.156...51M}, \citealt{mochnacki1972models}, \citealt{2024AJ....168..272P}).

In this study, ground-based multiband photometric observations were carried out for eight total-eclipsing W UMa-type contact binary systems with the goal of refining their orbital and physical parameter estimates. This study builds upon the efforts initiated by \cite{2025MNRAS.537.3160P, 2025AJ....170..214P, 2025PASP..137h4201P}, providing new observations and analyses of additional W UMa-type contact binaries within the framework of the BSN project\footnote{\url{https://bsnp.info/}}. The structure of this paper is as follows: Section 2 introduces the target systems and the observation processes used in this work. Section 3 addresses orbital period variations. Section 4 presents the light curve modeling results. Section 5 discusses the derivation of absolute stellar parameters. Finally, Section 6 provides a discussion and conclusion of the main findings and their implications.

\vspace{0.6cm}
\section{Target Systems and Observation Processes}
\label{sec2}
\subsection{Introducing and Characterizing}
We have analyzed eight eclipsing binary stars, including ASAS J165337+1542.7 (hereinafter J1653), CRTS J005742.9+203750 (hereinafter J0057), CRTS J015947.3+283711 (hereinafter J0159), NSVS 3794717 (hereinafter N3794), UCAC4 687-086841 (hereinafter U687), V3055 Cyg, V637 Peg , and ZTF J232250.75+355048.6 (hereinafter Z2322). These contact binary systems had not been studied in detail previously. Also, we had multiband photometric data available in the BSN project database for these systems, providing sufficient observational coverage for accurate analysis. Table \ref{Tab:systemsinfo} presents the specifications of the target systems from the Gaia DR3 database (\citealt{2023AA...674A..33G}), along with information on their discoverers. The catalogs in which these target systems were discovered are as follows: Robotic Optical Transient Search Experiment I (ROTSE-I; \citealt{2000AJ....119.1901A}), Catalina Surveys Data Release 1 (CSDR1; \citealt{2014ApJS..213....9D}), Northern Sky Variability Survey (NSVS; \citealt{2009AJ....138..466H}), Asteroid Terrestrial-impact Last Alert System (ATLAS; \citealt{2018ApJ...867..105T}), Zwicky Transient Facility (ZTF, \citealt{2020ApJS..249...18C}). The target systems are listed as contact binary systems in various catalogs and databases, such as the All-Sky Automated Survey for Supernovae (ASAS-SN; \citealt{2018MNRAS.477.3145J}), and the Variable Star Index (VSX\footnote{\url{https://vsx.aavso.org/}}).

\vspace{0.4cm}
\subsection{Ground-based Observations}
Ground-based photometric observations of the eight binary stars were obtained at the San Pedro Mártir Observatory in México, situated at $115^\circ27^{'}49^{''}$ W, $31^\circ02^{'}39^{''}$ N, at an elevation of 2830 m above sea level. The measurements were made with a 0.84 m Ritchey–Chrétien telescope ($f/15$) equipped with a Marconi-5 CCD camera from Spectral Instruments, incorporating an e2v CCD231-42 sensor with $15\times15,\mu\mathrm{m}^2$ pixels, a gain of $2.2\ e^-\ \mathrm{ADU}^{-1}$, and a readout noise of $3.6\ e^-$. Standard $B$, $V$, $R_c$, and $I_c$ filters were employed. Image processing followed standard procedures using IRAF, including bias removal and flat-field correction, in accordance with the methodology outlined by \cite{1986SPIE..627..733T}. Table \ref{Tab:observations} presents the observational details for each target system, including the observation date, filters used, and exposure times. Table \ref{Tab:stars} provides the general characteristics of the comparison and check stars used during the observation and data reduction processes. The information in Table \ref{Tab:stars} is from Gaia DR3 (\citealt{2023AA...674A..33G}).

\vspace{0.4cm}
\subsection{TESS Data}
Space-based time-series data for this study were obtained from the Transiting Exoplanet Survey Satellite (TESS). TESS employs four wide-field cameras to monitor different regions of the sky for exoplanet detection, observing each sector for approximately 27.4 days. We analyzed TESS photometric data for the eight target binary systems in the broad “TESS:T” band (600–1000 nm). The relevant sectors are listed in Table \ref{Tab:tess}. All observations were retrieved from the Mikulski Archive for Space Telescopes (MAST) and processed using the Lightkurve package, with detrending procedures consistent with the Science Processing Operations Center (SPOC) pipeline.

\begin{table*}
\renewcommand\arraystretch{1.2}
\caption{Specifications of the target systems: coordinates from Gaia DR3, V–R values from the TIC, and their discoverers.}
\centering
\begin{center}
\footnotesize
\begin{tabular}{c c c c c c c}
\hline
System & RA$.^\circ$(J2000) & Dec$.^\circ$(J2000) & $d$(pc) & $T_{Gaia}$(K) & $V-R$(mag.) & Discover\\
\hline
ASAS J165337+1542.7 (J1653)	& 253.4041846 &	15.7104837 & 434(6) & 5111(36) & 0.484 & ROTSE-I\\
CRTS J005742.9+203750 (J0057) & 14.4290608 & 20.6305464 & 444(4) & 5099(15) & 0.462 & CSDR1\\
CRTS J015947.3+283711 (J0159) & 29.9474594 & 28.6201850 & 550(7) & 5100(34) & 0.471 & CSDR1\\
NSVS 3794717 (N3794) & 13.5394471 & 38.8952815 & 528(4) & 5633(10) & 0.352 & NSVS\\
UCAC4 687-086841 (U687)	& 312.3845578 &	47.3112635 & 564(5) & 5085(24)  & 0.468 & ATLAS\\
V3055 Cyg & 308.9617905 &	52.7038444 & 749(5) & 6052(68) & 0.429 & \cite{2009PZP.....9...23D}\\
V637 Peg & 335.4724193 & 28.0463187 & 316(2) & 4904(37) & 0.519 & NSVS\\
ZTF J232250.75+355048.6 (Z2322)	& 350.7114725 &	35.8468252 & 226(4) & - & 0.517 & ZTF\\
\hline
\end{tabular}
\end{center}
\label{Tab:systemsinfo}
\end{table*}

\begin{table*}
\renewcommand\arraystretch{1.2}
\caption{Specifications of the ground-based observations.}
\centering
\begin{center}
\footnotesize
\begin{tabular}{c c c}
\hline
System & Observation(s) Date & Filter and Exposure time(s)\\
\hline
J1653	& 2024 (May 27) & $B(90)$, $V(50)$, $R_c(35)$, $I_c(30)$\\
J0057	& 2024 (October 9) & $B(90)$, $V(50)$, $R_c(35)$, $I_c(30)$\\
J0159	& 2024 (October 14) & $B(90)$, $V(50)$, $R_c(35)$, $I_c(30)$\\
N3794	& 2024 (October 4) & $B(90)$, $V(50)$, $R_c(35)$, $I_c(30)$\\
U687 & 2024 (August 3) & $B(90)$, $V(50)$, $R_c(35)$, $I_c(30)$\\
V3055 Cyg & 2024 (August 11,13) & $B(21)$, $V(11)$, $R_c(9)$, $I_c(7)$\\
V637 Peg & 2024 (August 16) & $B(90)$, $V(50)$, $R_c(35)$, $I_c(30)$\\
Z2322	& 2024 (September 19) & $B(90)$, $V(50)$, $R_c(35)$, $I_c(30)$\\
\hline
\end{tabular}
\end{center}
\label{Tab:observations}
\end{table*}

\begin{table*}
\renewcommand\arraystretch{1.2}
\caption{The comparison and check stars in the ground-based observations. Coordinates are taken from the Gaia DR3, while the $V-R$ values come from the TIC.}
\centering
\begin{center}
\footnotesize
\begin{tabular}{c c c c c c}
\hline
System & Star Type & Star Name & RA$.^\circ$(J2000) & DEC$.^\circ$(J2000) & $V-R$(mag.)\\
\hline
J1653	&	Comparison	& 2MASSJ16532375+1540551 & 253.34891 &		15.68196 & 0.499\\
J1653	&	Check	& 2MASSJ16534970+1544396 & 253.45703 & 15.74437 & 0.543\\
J0057	&	Comparison	& 2MASSJ00574261+2039520 & 14.42765 &		20.66446 & 0.426\\
J0057	&	Check	&	2MASSJ00580307+2040090	& 14.51280 &		20.66908 & 0.437\\
J0159	&	Comparison	& 2MASSJ01594552+2834065 & 29.93968 &		28.56845 & 0.513\\
J0159	&	Check	&	2MASSJ01595184+2833230	& 29.96602 &		28.55637 & 0.444\\
N3794	&	Comparison	& 2MASSJ00540020+3853200 & 13.50091 &		38.88892 & 0.366\\
N3794	&	Check	& 2MASSJ00534896+3857353 & 13.45402 & 38.95979 & 0.373\\
U687	&	Comparison	& 2MASSJ20491628+4717079 & 312.31785 &		47.28547 & 0.442\\
U687	&	Check	& 2MASSJ20493614+4717193 & 312.40068 & 47.28867 & 0.408\\
V3055 Cyg	&	Comparison	& 2MASSJ20354993+5245072 & 308.95796 & 52.75189 & 0.309\\
V3055 Cyg	&	Check	& 2MASSJ20354199+5244172 & 308.92501 &		52.73811 & 0.152\\
V637 Peg	&	Comparison	& 2MASSJ22213884+2805262 & 335.41184 & 28.09069 & 0.500\\
V637 Peg	&	Check	& 2MASSJ22220180+2806152 & 335.50758 &		28.10423 & 0.510\\
Z2322	&	Comparison	& 2MASSJ23222399+3552418 & 350.60008 &		35.87824 & 0.490\\
Z2322	&	Check	& 2MASSJ23230008+3548138 & 350.75040 & 35.80380 & 0.561\\
\hline
\end{tabular}
\end{center}
\label{Tab:stars}
\end{table*}

\begin{table*}
\renewcommand\arraystretch{1.2}
\caption{Specifications of the TESS data used in this study.}
\centering
\begin{center}
\footnotesize
\begin{tabular}{c c c c c c}
\hline
System & TIC & TESS Sector & Exposure Length(s) & Observation Year\\
\hline
J1653	&	353179421	&	25,79	&	1800,200	&	2020,2024	\\
J0057	&	436554808	&	17	&	1800	&	2019	\\
J0159	&	28358711	&	17,58	&	1800,200	&	2020,2023	\\
N3794	&	283569299	&	17,57	&	1800,200	&	2020,2023	\\
U687	&	353335581	&	15	&	1800	&	2022	\\
V3055 Cyg	&	303755872	&	15,16,41,55,56,75,76,82	&	1800,1800,600,600,200,200,200,200	&	2019,2019,2021,2022,2022,2024,2024,2024	\\
V637 Peg	&	20637181	&	56	&	200	&	2023	\\
Z2322	&	21283546	&	57	&	200	&	2022	\\
\hline
\end{tabular}
\end{center}
\label{Tab:tess}
\end{table*}

\vspace{0.6cm}
\section{Orbital Period Variations}
\label{sec3}
To conduct a detailed analysis of the targets, it was essential to compile a comprehensive set of eclipse timing measurements from both archival photometric surveys and our own observations. These timing measurements were then used to investigate orbital period variations across the entire systems. As part of this effort, we queried the VarAstro\footnote{\url{http://var.astro.cz}}, and VSX databases for published minima times. In addition, eclipse timings were extracted directly from the available TESS time series data. For the 2-minute cadence TESS data, individual eclipse timings were derived straightforwardly from the light curves. However, due to the lower temporal resolution of the 30-minute cadence data, we first applied a phase-folding procedure following the method described by \cite{Li_2020} to combine observations across multiple cycles. After phase alignment, eclipse timings were extracted from the folded light curves.

Eclipse minima extracted from our multiband photometric observations contributed to improving the temporal coverage and supported the overall reliability of the period variation analysis, complementing the minima gathered from archival surveys.

Ensuring consistency in timing measurements across all datasets required converting Heliocentric Julian Dates ($HJD$) into Barycentric Julian Dates in Barycentric Dynamical Time ($BJD_{TDB}$)(\citealt{2010PASP..122..935E}). This transformation was carried out using the online tool\footnote{\url{https://astroutils.astronomy.osu.edu/time/hjd2bjd.html}} due to the presence of both time formats in our data.

The extraction of eclipse minima times was initially performed using the classical technique developed by \citet{kw1956}, a method long established as standard in eclipsing binary light curve analysis. However, prior studies have shown that this approach often leads to underestimated timing uncertainties \citep[e.g.,][]{2012AN....333..754P, 2018MNRAS.480.4557L}, and may not perform well for light curves that are asymmetric or incomplete \citep{mikulavsek2013kwee}. To improve the reliability of the timing measurements, we applied a method based on profile fitting, inspired by the work of \citet{2021AstL...47..402P}. In this approach, we modeled the eclipse dips using Gaussian or Cauchy functions, focusing on segments of the light curves around the minima. We estimated the parameter uncertainties, including those of the mid-eclipse times, using Markov Chain Monte Carlo (MCMC) simulations, implemented via the \texttt{emcee} Python package \citep{2013PASP..125..306F}. This methodology provided more robust and realistic uncertainty estimates, particularly for shallow and asymmetric eclipse minima.

\begin{figure*}
\centering
\includegraphics[width=0.78\textwidth]{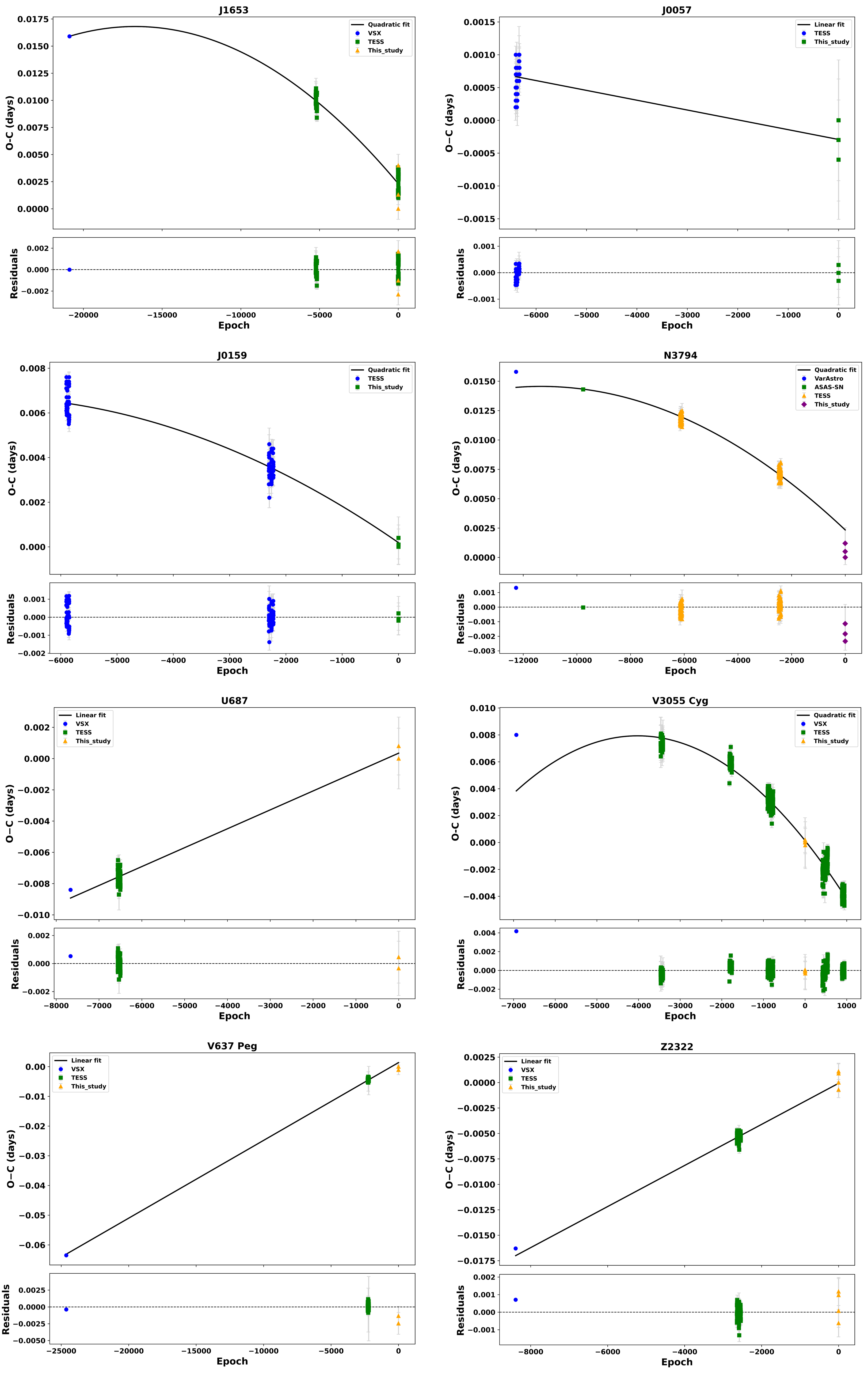}
\caption{O-C diagrams for the eight target systems, with corresponding residuals shown in the bottom panel.}
\label{Fig:oc}
\end{figure*}

The eclipse timings from our observations are listed in Table \ref{Tab:min}. A machine-readable version of the complete compiled eclipse timing data for the target binary systems is available online. Subsequently, the O-C values were calculated based on the reference ephemeris:
\begin{equation}\label{eq:OC}
BJD=BJD_0+P\times E,
\end{equation}
\noindent where $BJD$ refers to the observed eclipse timing, $BJD_0$, listed in the second column of Table \ref{Tab:ephemeris}, indicates the reference epoch of the primary eclipse, and $P$, found in the third column of the same table, represents the orbital period. The computed epochs together with the O–C values are presented in Table \ref{Tab:min} and are also accessible in a machine-readable format online. Corresponding O–C diagrams are illustrated in Figure \ref{Fig:oc}.

Based on the available eclipse minima timings, four of the systems exhibited linear trends, while another four targets showed parabolic variations. For those systems displaying parabolic behavior, the following equation was used for fitting the O-C diagrams:
\begin{equation}\label{parabolic}
O-C=\Delta{T_0}+\Delta{P_0}\times E+\frac{\beta}{2}{E^2}.
\end{equation}

Updated ephemerides for the targets are provided in Table \ref{Tab:ephemeris}. Table \ref{Tab:mass-transfer} lists the fitted parameters and the calculated mass transfer rates for the four systems exhibiting parabolic period variations.

\renewcommand\arraystretch{1.2}
\begin{table*}
\caption{The times of minima extracted from our ground-based observations.}
\centering
\small
\begin{tabular}{c c c c c}
\hline
System & Min.($BJD_{TDB}$) & Error & Epoch & O-C\\ 
\hline
J1653	&	2460457.6787	&	0.0009	&	-0.5	&	0.0013	\\
	&	2460457.8178	&	0.0010	&	0.0	&	0	\\
	&	2460457.9623	&	0.0010	&	0.5	&	0.0040	\\
J0057	&	2460592.7024	&	0.0009	&	-0.5	&	-0.0003	\\
	&	2460592.8454	&	0.0009	&	0.0	&	0	\\
	&	2460592.9874	&	0.0009	&	0.5	&	-0.0006	\\
J0159	&	2460597.6655	&	0.0008	&	0.0	&	0	\\
	&	2460597.8212	&	0.0009	&	0.5	&	0.0004	\\
	&	2460597.9762	&	0.0009	&	1.0	&	0.0001	\\
N3794	&	2460587.6953	&	0.0006	&	-1.0	&	0.0005	\\
	&	2460587.8440	&	0.0013	&	-0.5	&	0.0012	\\
	&	2460587.9908	&	0.0006	&	0.0	&	0	\\
U687	&	2460525.8090	&	0.0019	&	0.0	&	0	\\
	&	2460525.9481	&	0.0019	&	0.5	&	0.0008	\\
V3055 Cyg	&	2460167.7352	&	0.0018	&	0.0	&	0	\\
	&	2460167.9414	&	0.0009	&	0.5	&	0.0002	\\
	&	2460169.5889	&	0.0017	&	4.5	&	-0.0002	\\
	&	2460169.7952	&	0.0009	&	5.0	&	0.0001	\\
V637 Peg	&	2460538.7905	&	0.0014	&	0.0	&	0	\\
	&	2460538.9453	&	0.0016	&	0.5	&	-0.0011	\\
Z2322	&	2460572.7424	&	0.0008	&	-0.5	&	-0.0007	\\
	&	2460572.7442	&	0.0008	&	-0.5	&	0.0011	\\
	&	2460572.8792	&	0.0009	&	0.0	&	0	\\
	&	2460573.0163	&	0.0009	&	0.5	&	0.0009	\\
\hline
\end{tabular}
\label{Tab:min}
\end{table*}

\renewcommand\arraystretch{1.2}
\begin{table*}
\caption{Reference and new ephemeris of the eight systems. The reference times of minimum ($t_0$) were obtained from our observations in this study.}
\centering
\small
\begin{tabular}{c|cc|cc}
\hline
System& \multicolumn{2}{c|}{Reference ephemeris}& \multicolumn{2}{c}{New ephemeris}\\ 
&$t_0(BJD_{TDB})$&Period(day)/Source& Corrected $t_0(BJD_{TDB})$&New Period(day)\\ 
\hline
J1653 & 2460457.8178(10) & 0.2809786/ASAS-SN & 2460457.8202(2) & 0.28097687(5)\\
J0057 & 2460592.8454(9) & 0.2852965/ASAS-SN & 2460592.8451(1) & 0.28529635(2)\\
J0159 & 2460597.6655(8) & 0.3105738/ASAS-SN & 2460597.6657(3) & 0.3105721(2)\\
N3794 & 2460587.9908(6) & 0.296089/ASAS-SN & 2460587.9931(2) & 0.29608684(8)\\
U687 & 2460525.8090(19) & 0.2765148/VSX & 2460525.8093(3) & 0.27651601(5)\\
V3055 Cyg & 2460167.7352(18) & 0.41198165/VSX & 2460167.7354(1) & 0.41197778(4)\\
V637 Peg & 2460538.7905(4) & 0.311791/VSX & 2460538.7919(2) & 0.31179362(3)\\
Z2322 & 2460572.8792(9) & 0.2722212/VSX & 2460572.8791(2) & 0.27222322(6)\\
\hline
\end{tabular}
\label{Tab:ephemeris}
\end{table*}

\renewcommand\arraystretch{1.2}
\begin{table*}
\caption{The O-C fitting coefficients and mass transfer rate.}
\centering
\small
\begin{tabular}{ccccccccc}
\hline
Parameter& $\Delta{T_0}$& Error& $\Delta{P_0}$& Error& $\beta$&Error & $dM_1/dt$&Error\\ 
&$(\times {10^{-4}} \text{d})$&& $(\times {10^{-7}} \text{d})$&& $(\times {10^{-7}} \text{d}$ $ {\text{yr}^{-1}})$& & $(\times {10^{-7}} M_\odot$ $ {\text{yr}^{-1}})$&\\ 
\hline
J1653 & 23.1 & 1.2 & -17.4 & 0.5 & -1.4 & 0.1 & 0.6 & 0.1\\
J0159 & 1.9 & 2.9 & -17.5 & 1.9 & -2.7 & 0.6 & 2.5 & 0.7\\
N3794 & 23.4 & 1.7 & -21.6 & 0.8 & -2.4 & 0.2 & 2.0 & 0.2\\
V3055 Cyg & 1.4 & 0.4 & -38.8 & 0.4 & -8.6 & 0.2 & -1.9 & 0.3\\
\hline
\end{tabular}
\label{Tab:mass-transfer}
\end{table*}

\vspace{0.6cm}
\section{Light Curve Solutions}
\label{sec4}
We analyzed the light curves of the eight systems based on our ground-based multiband observations, combined with the time-series data from the latest available TESS sector. The light curve modeling process began with the conversion of time data into orbital phase, using the updated ephemeris provided in Table \ref{Tab:ephemeris}. Light curve analysis of the target binary stars was carried out using version 2.4.9 of the PHysics Of Eclipsing BinariEs (PHOEBE) Python code \citep{2016ApJS..227...29P, 2020ApJS..250...34C}. Based on the morphology of the observed light curves, classifications from existing catalogs, and the systems' short orbital periods, the contact configuration was selected as the most appropriate modeling approach.

The modeling was carried out under a set of standard assumptions commonly adopted for close binary systems. The gravity-darkening coefficients were assumed to be $g_1 = g_2 = 0.32$ \citep{1967ZA.....65...89L}, and the bolometric albedos were set to $A_1 = A_2 = 0.5$ \citep{1969AcA....19..245R}, values that are typical for stars with convective outer layers. The stellar atmosphere model used in the analysis followed the framework presented by \cite{2004AA...419..725C}. Limb-darkening coefficients were not fixed but instead treated as free parameters within PHOEBE, allowing for improved fitting flexibility and better alignment with the observed light curves.

Initial estimates for several key parameters were determined to begin the light curve modeling process. For seven of the target systems, the effective temperature ($T$) was adopted from the Gaia DR3 database and assigned to the hotter component of each binary, based on the relative depths of the primary and secondary eclipses. The effective temperature of the cooler star was then estimated by measuring the depth difference between the two minima in the observed light curves. For the system Z2322, no temperature values were available in either the Gaia DR3 database or the TIC; therefore, the effective temperature of the hotter component was estimated using an empirical relationship between orbital period and stellar temperature, as described in detail in Section 6, Part B.

\begin{figure*}
\centering
\includegraphics[width=0.9\textwidth]{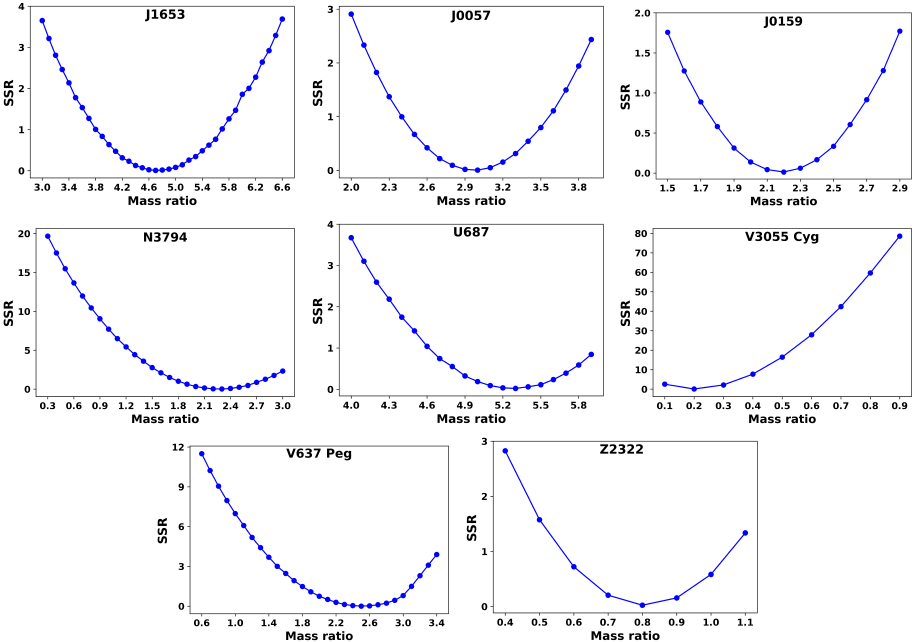}
\caption{The sum of squared residuals as a function of mass ratio.}
\label{Fig:q-diagrams}
\end{figure*}

The initial mass ratio ($q$) of the systems was estimated using the $q$-search method (\citealt{2005ApSS.296..221T}). A broad mass ratio range from 0.1 to 20 was initially explored for each target system. This range was then narrowed to refine the mass ratio estimate by minimizing the sum of squared residuals between the observed and synthetic light curves. As shown in Figure \ref{Fig:q-diagrams}, each $q$-search curve displays a distinct minimum in the sum of squared residuals, indicating the best-fit value of $q$. Previous studies, such as \citet{2024AJ....168..272P}, have shown that the $q$-search method yields more reliable mass ratio estimates for systems with total eclipses than for those exhibiting partial eclipses.

Asymmetry in light curve maxima is common in contact binary systems and is observed in four of our targets (Table \ref{Tab:lc-analysis}). This is typically attributed to starspots caused by magnetic activity, a phenomenon known as the O'Connell effect \citep{1951PRCO....2...85O}. Although magnetic activity and single starspots are commonly considered in modeling, alternative mechanisms have also been proposed to more fully explain the observed asymmetries \citep{1990ApJ...355..271Z, 2003ChJAA...3..142L}.

The light curves of the target systems were modeled without incorporating $l_3$. No evidence of contamination from nearby stars was detected, nor is there current support for the presence of a third light component. However, the possibility of an $l_3$ contribution cannot be definitively excluded.

Using multiband photometric data ($BVR_cI_c$) and initial parameter estimates, we obtained a satisfactory theoretical fit. The optimization tool of the PHOEBE code was then employed to refine the light curve solution, yielding more precise values for the effective temperatures, mass ratio, fillout factor, and orbital inclination.

PHOEBE's built-in modeling and optimization tools do not directly provide uncertainty estimates for fitted parameters. To address this, we utilized the BSN application version 1.0 \citep{paki2025bsn}, a Windows-compatible program specifically designed to accelerate MCMC fitting processes. BSN application achieves a computational speed exceeding that of PHOEBE by over 40 times when generating synthetic light curves. This enhancement is mainly attributed to BSN application's optimized architecture and incorporation of advanced computational libraries, while maintaining analysis techniques consistent with standard binary star modeling frameworks. For the MCMC analysis, we configured BSN application with 28 walkers and 2000 iterations to sample five key parameters ($T_{1,2}$, $q$, $f$, and $i$). Uncertainties for the parameters were estimated based on the MCMC simulation results, and the average upper and lower uncertainty limits for each parameter are presented in Table \ref{Tab:lc-analysis}. Importantly, the parameter values and synthetic light curves obtained through BSN application closely matched those generated by PHOEBE, confirming the consistency of the results.

\begin{figure*}
\centering
\includegraphics[width=0.48\textwidth]{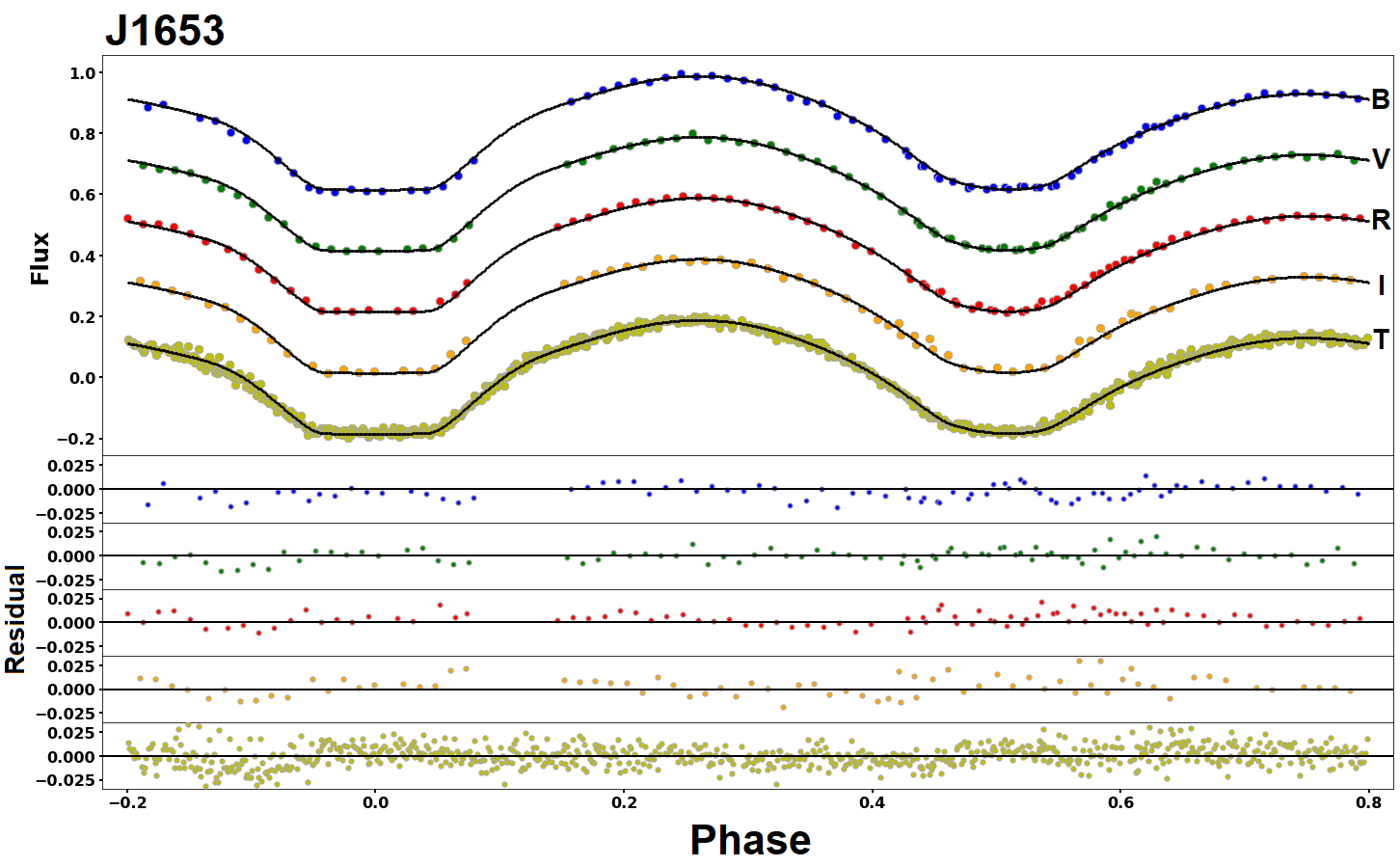}
\includegraphics[width=0.48\textwidth]{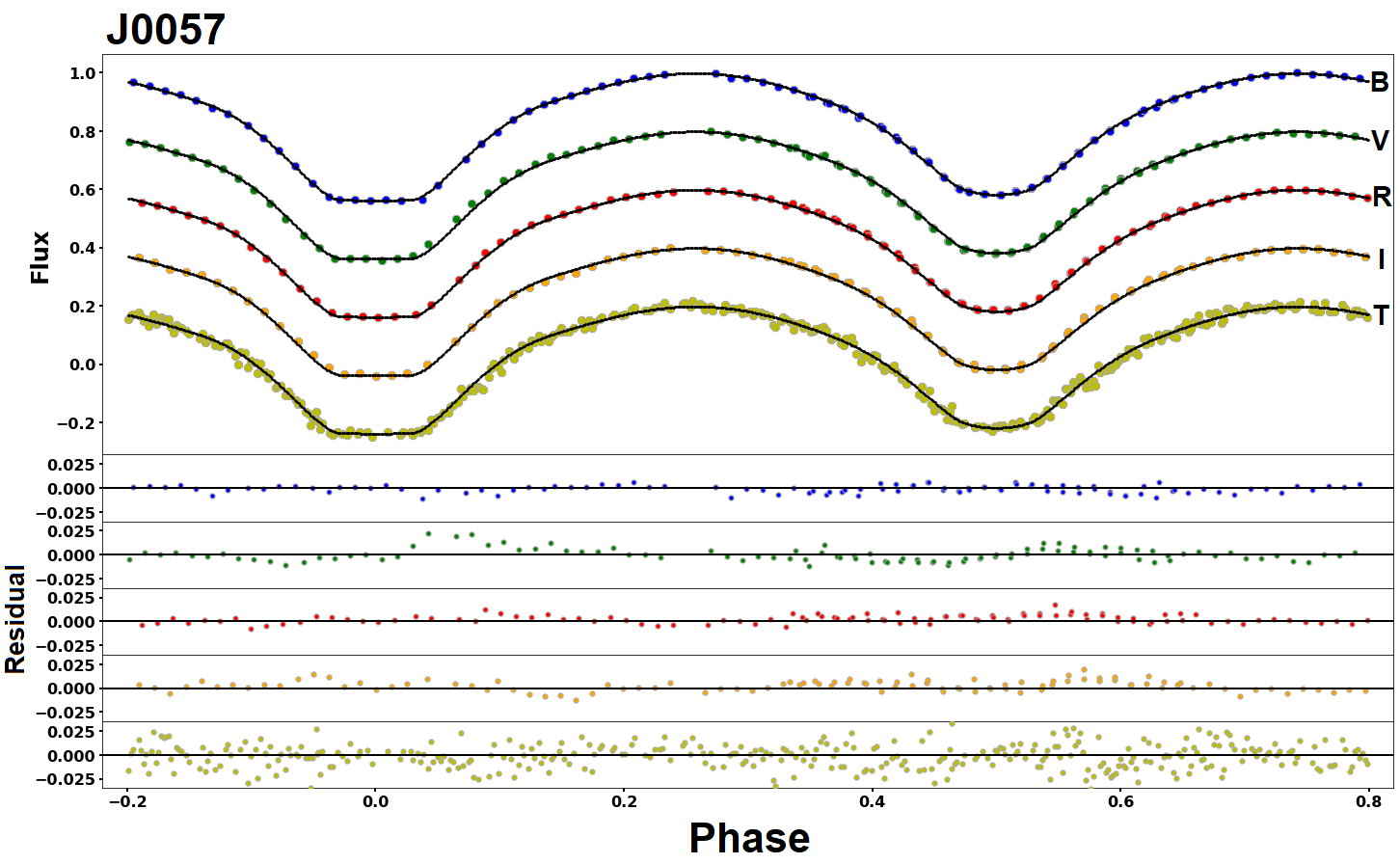}
\includegraphics[width=0.48\textwidth]{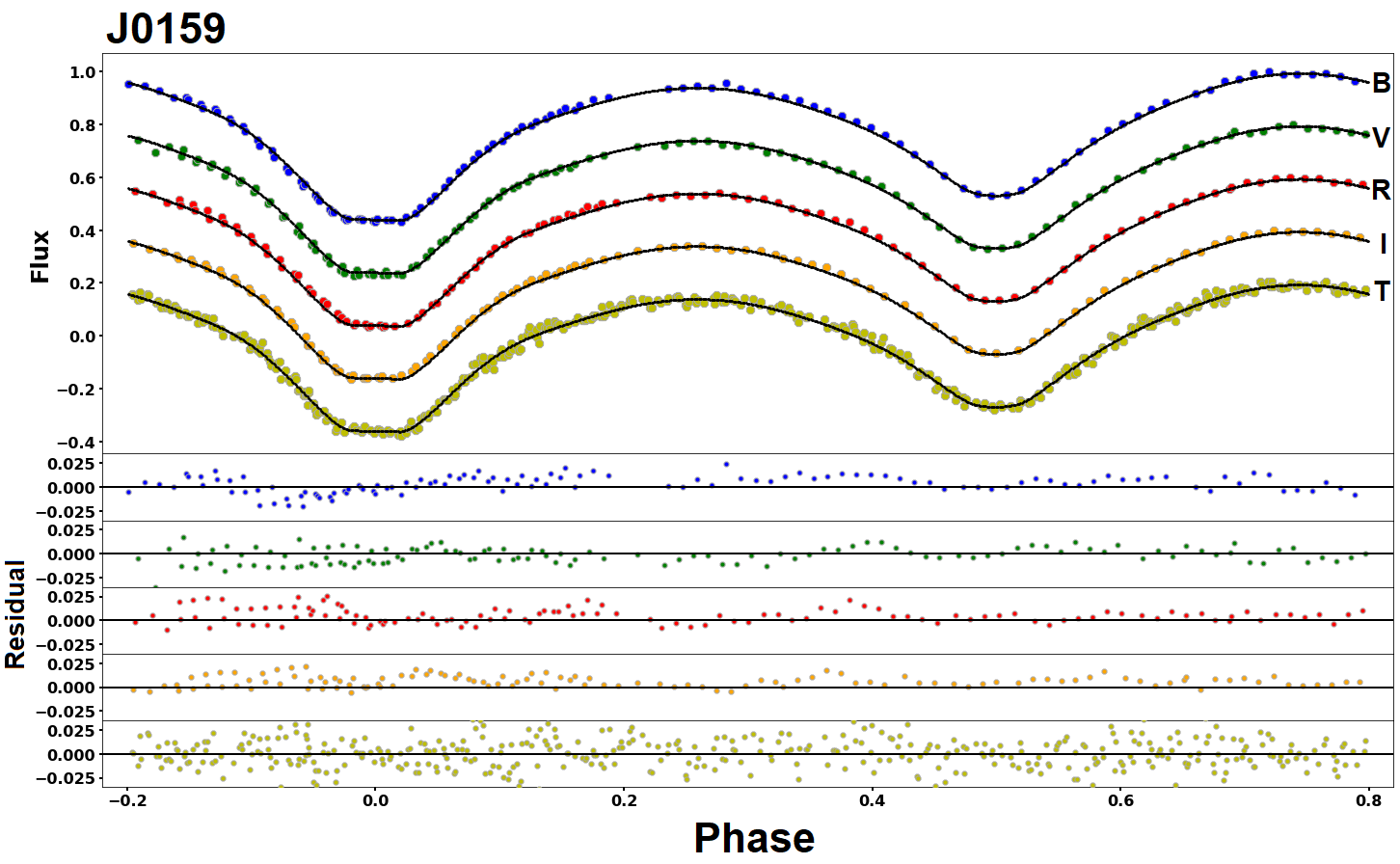}
\includegraphics[width=0.48\textwidth]{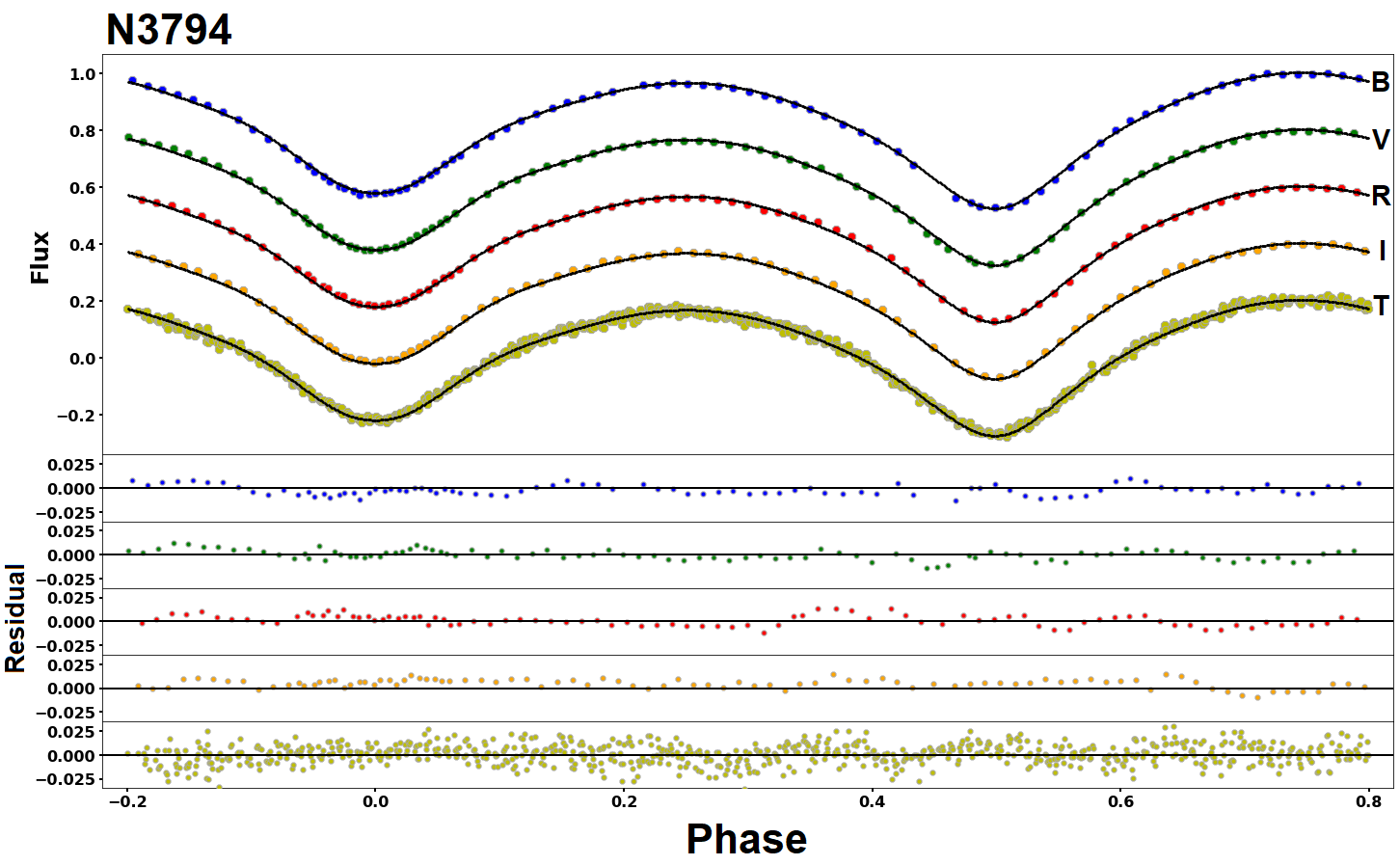}
\includegraphics[width=0.48\textwidth]{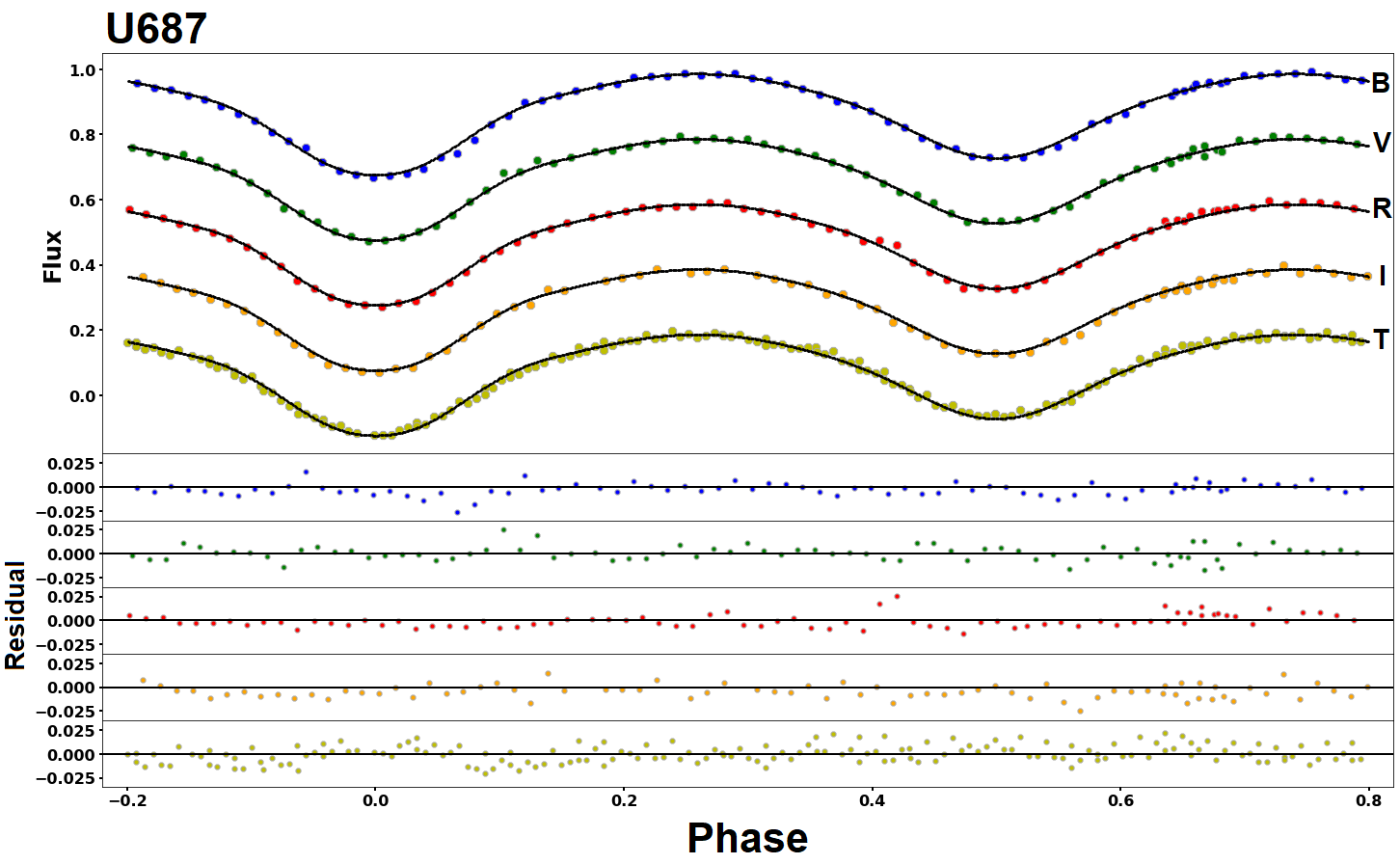}
\includegraphics[width=0.48\textwidth]{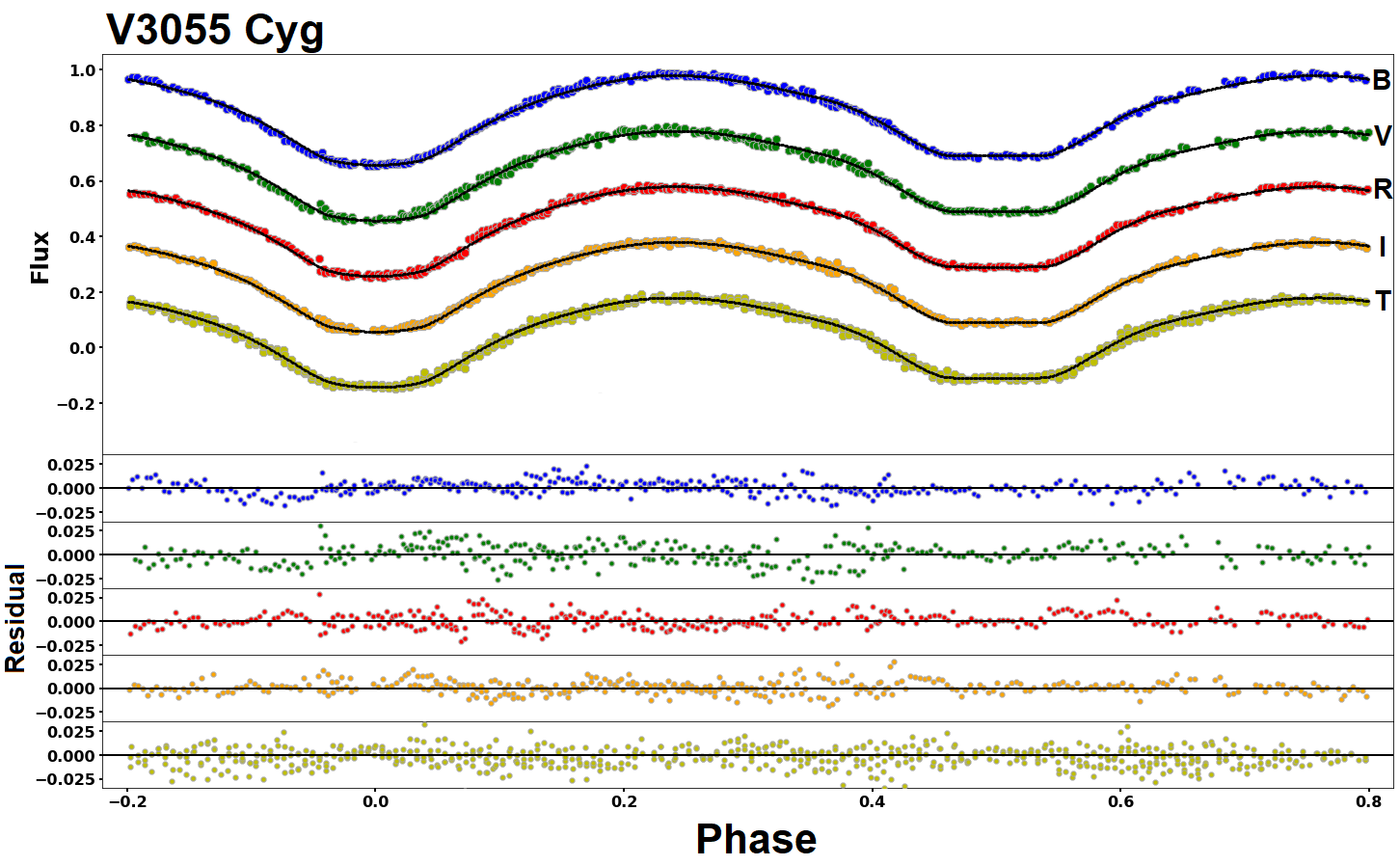}
\includegraphics[width=0.48\textwidth]{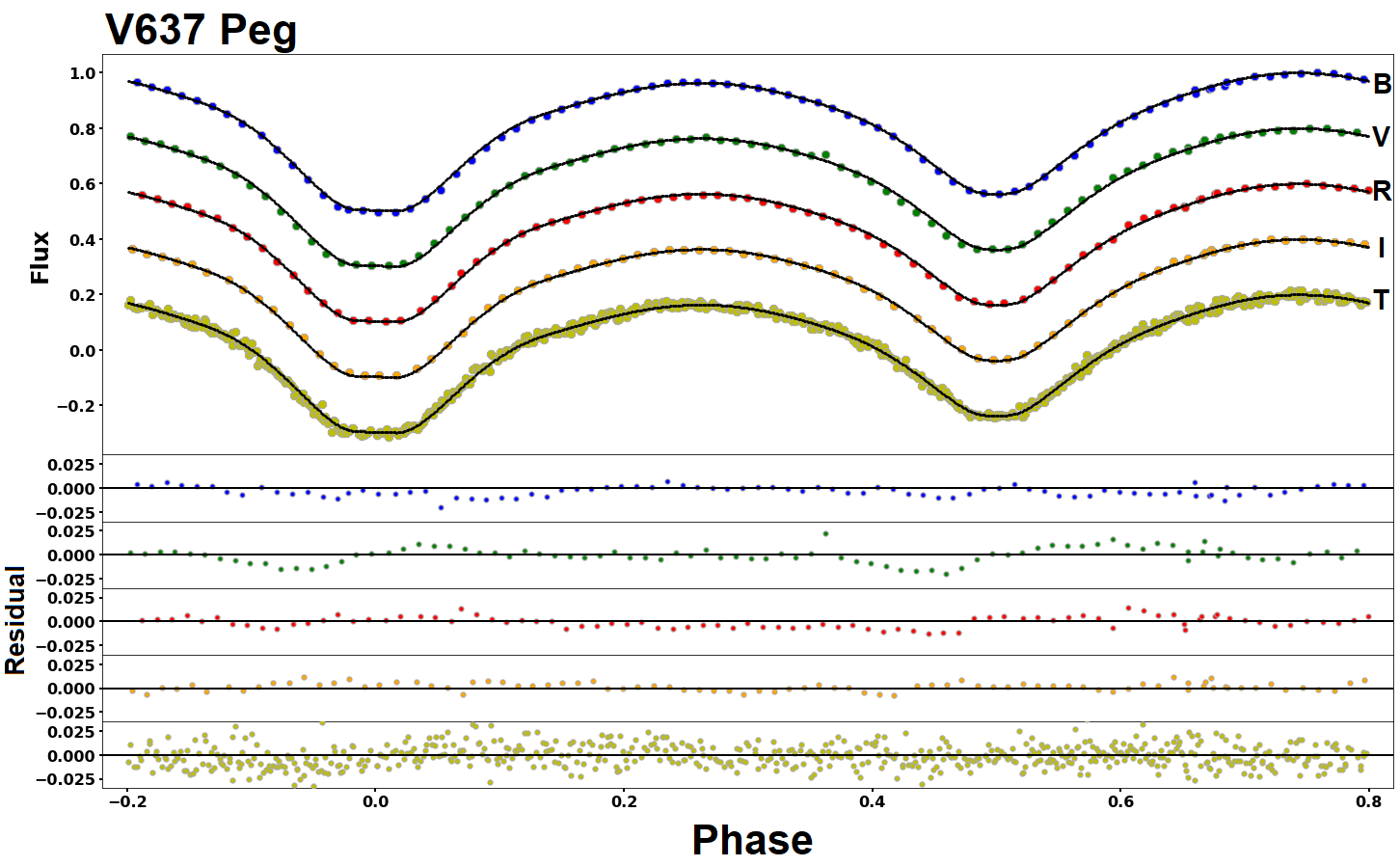}
\includegraphics[width=0.48\textwidth]{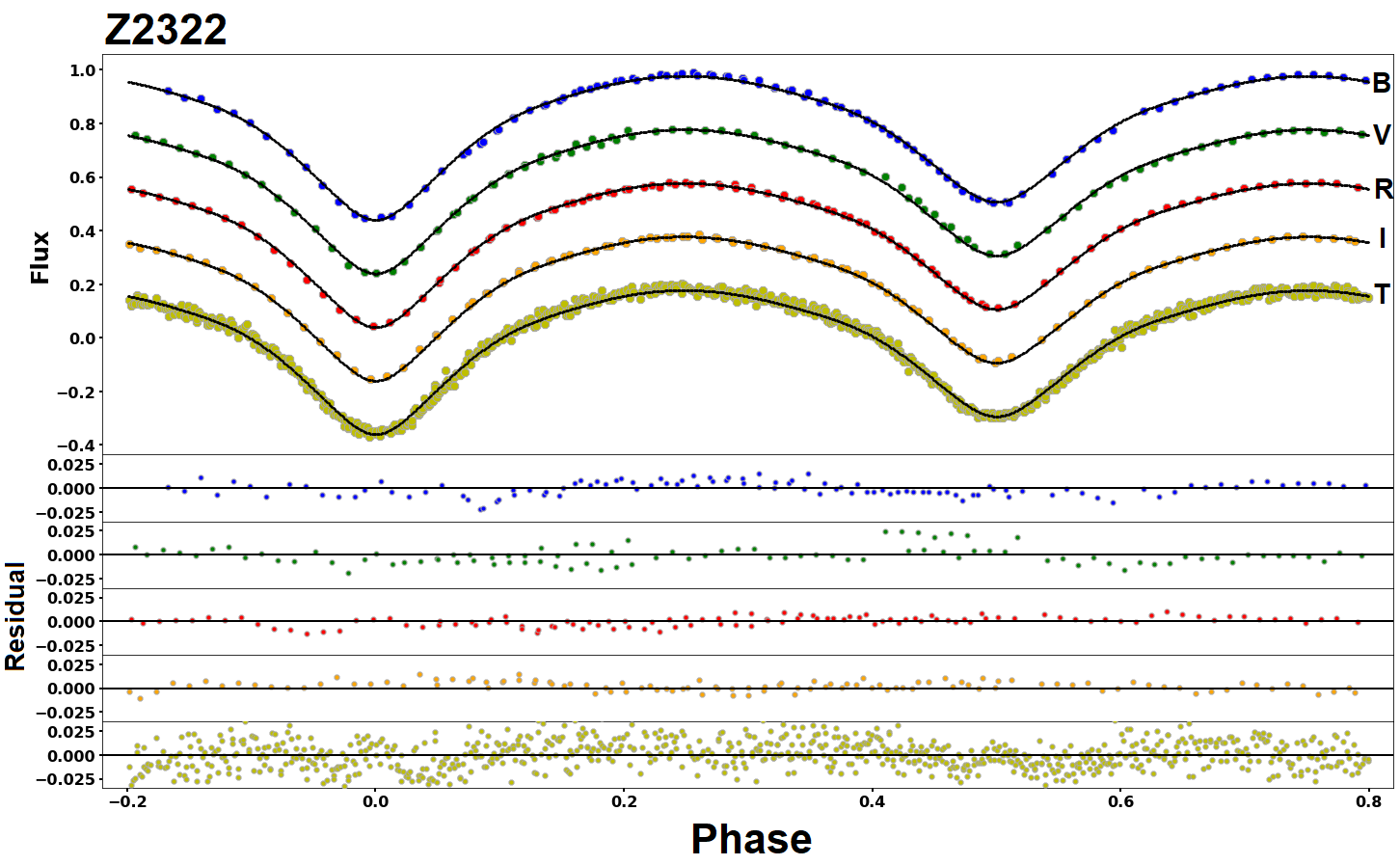}
\caption{The colored dots represent the observed light curves of the systems in different filters, and the synthetic light curves, generated using the light curve solutions, are also shown. Residuals are shown at the bottom of each panel.}
\label{Fig:lc}
\end{figure*}

\begin{figure*}
\centering
\includegraphics[width=0.9\textwidth]{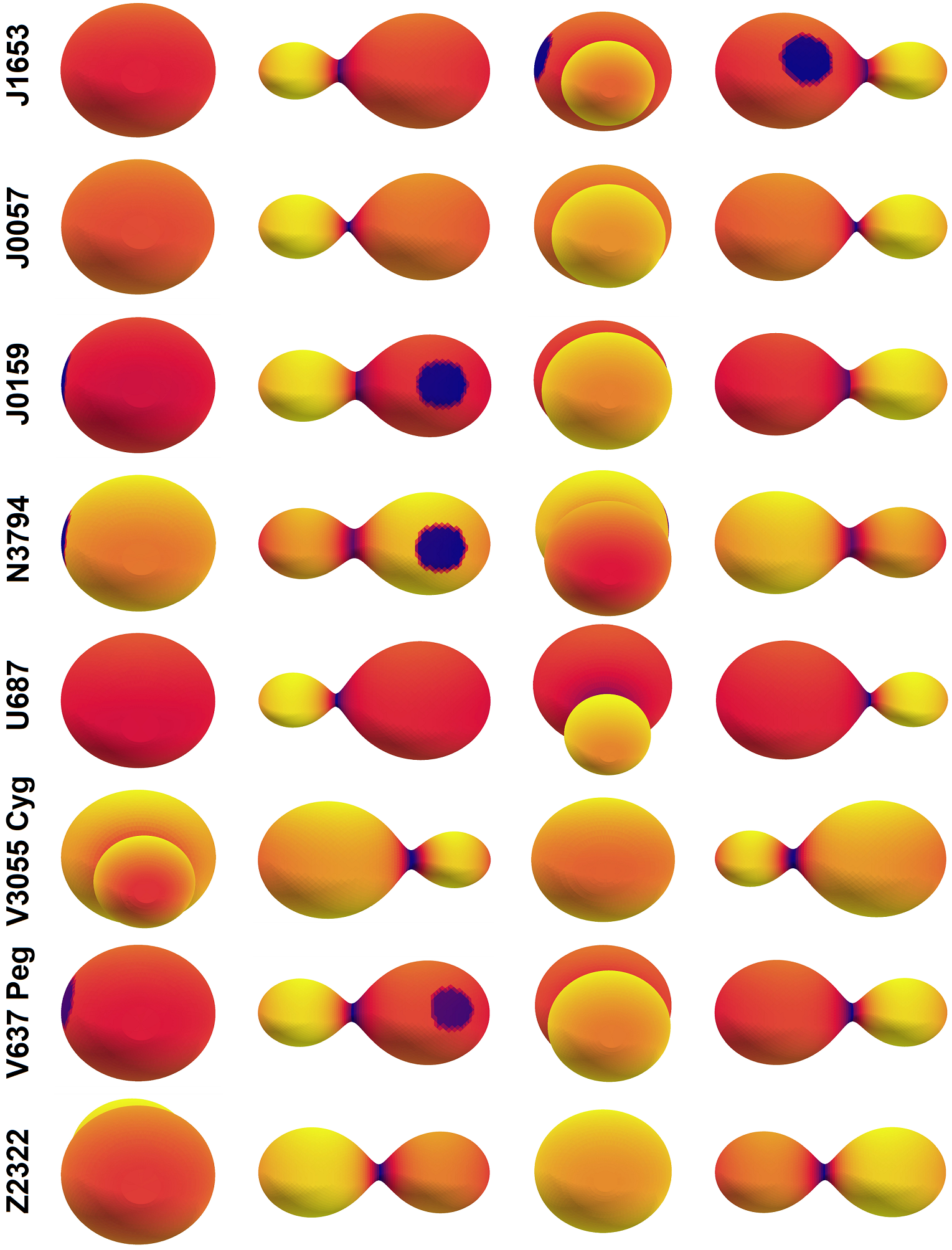}
\caption{Three-dimensional representations of the target binary stars shown at orbital phases 0, 0.25, 0.5, and 0.75, respectively.}
\label{Fig:3d}
\end{figure*}

The final results of the light curve analysis, including starspot characteristics, are summarized in Table \ref{Tab:lc-analysis}. The observed and synthetic light curves in different filters are displayed in Figure \ref{Fig:lc}. Three-dimensional representations of the binary systems, constructed from the final model parameters, are presented in Figure \ref{Fig:3d}.

\begin{table*}
\renewcommand\arraystretch{1.5}
\caption{Light curve solutions of the target binary stars.}
\centering
\begin{center}
\footnotesize
\begin{tabular}{c c c c c c c c c}
\hline
Parameter & J1653 & J0057 & J0159 & N3794 & U687 & V3055 Cyg & V637 Peg & Z2322\\
\hline
$T_{1}$ (K) & 5022(46) & 5176(35) &	5208(37) & 5559(41) & 5349(29) & 6133(48) & 4975(28) & 5405(36)\\
$T_{2}$ (K) & 4768(51) & 4952(33) &	4818(36) & 5675(44) & 4938(32) & 6130(47) & 4720(25) & 5222(34)\\
$q=M_2/M_1$ & 4.739(29) & 2.971(37) & 2.207(18) & 2.277(22) & 5.297(35) &	0.200(8) & 2.515(21) & 0.809(10)\\
$i^{\circ}$ & 85.90(82) & 86.58(61) & 86.48(72) & 80.45(53) & 72.50(51) &	82.36(78) &	83.43(82) & 83.34(66)\\
$f$ & 0.299(11) & 0.105(3) & 0.288(9) &	0.321(8) & 0.103(7) & 0.213(10) &	0.130(7) & 0.072(3)\\
$\Omega_1=\Omega_2$ & 8.650(306) & 6.513(279) &	5.368(174) & 5.445(194) & 9.463(320) & 2.206(89) & 5.886(289) & 3.400(127)\\
$l_1/l_{tot}$($V$) & 0.259(3) &	0.326(4) & 0.443(4) & 0.306(5) & 0.237(4) & 0.807(5) & 0.377(4) & 0.593(5)\\
$l_2/l_{tot}$($V$) & 0.741(3) &	0.674(4) & 0.557(4) & 0.694(5) & 0.763(4) & 0.193(2) & 0.623(4) & 0.407(5)\\
$r_{(mean)1}$ &	0.270(16) &	0.296(7) & 0.334(21) & 0.334(24) & 0.252(5) & 0.533(9) & 0.311(9) & 0.405(6)\\
$r_{(mean)2}$ &	0.532(13) &	0.484(6) & 0.471(20) & 0.476(22) & 0.533(5) & 0.261(11) & 0.471(8) & 0.368(6)\\
\hline
$Col.^\circ$(spot) & 104 & - & 95 &	94 & - & - & 100 & -\\
$Long.^\circ$(spot) & 289 &	- &	96 & 71 & - & - & 107 & -\\
$Radius^\circ$(spot) & 20 &	- &	22 & 18 & - & - & 19 & -\\
$T_{spot}/T_{star}$ & 0.87 & - & 0.87 &	0.89 & - & - & 0.91 & -\\
Component &	Secondary &	- &	Secondary &	Secondary &	- &	- &	Secondary & -\\
\hline
\end{tabular}
\end{center}
\label{Tab:lc-analysis}
\end{table*}

\vspace{0.6cm}
\section{Absolute Parameters Estimations}
\label{sec5}
We utilized parallaxes from Gaia DR3 to determine the absolute parameters of the target binary systems, a method particularly advantageous in cases where only photometric data are available. To ensure the reliability of Gaia DR3 parallaxes for our analysis, we first estimated the interstellar extinction ($A_V$) affecting each system. According to the criteria outlined by \citet{2024PASP..136b4201P}, extinction values greater than approximately 0.4 can compromise the accuracy of this approach. Employing the three-dimensional dust maps from \citet{2019ApJ...887...93G}, we calculated $A_V$ values for each system and identified seven targets that fall within the acceptable range (Table \ref{Tab:absolute}). In contrast, system U687 exhibited a significantly higher extinction of $A_V=0.786(3)$, and was therefore excluded from absolute parameter estimation using Gaia DR3 parallaxes, due to exceeding the recommended extinction threshold for this method.

The absolute magnitude ($M_V$) of each system was calculated using the observed maximum brightness ($V_{\text{max}}$), the distances derived from Gaia DR3 parallaxes, and the corresponding interstellar extinction values. The $V_{\text{max}}$ measurements utilized in these calculations were obtained from our observational dataset, as listed in Table \ref{Tab:absolute}. Following this, the absolute magnitudes of the individual components, $M_{V1}$ and $M_{V2}$, were determined based on the luminosity ratios ($l_{1,2}/l_{\text{tot}}$) derived from the $V$-band light curve analysis. Bolometric magnitudes ($M_{\text{bol},1}$ and $M_{\text{bol},2}$) were then calculated by applying bolometric corrections ($BC_1$ and $BC_2$) according to the calibration provided by \citet{1996ApJ...469..355F}. Stellar luminosities ($L$) were estimated using the relationship between absolute bolometric magnitude and luminosity, adopting a solar bolometric magnitude of $M_{\text{bol}\odot} = 4.73$ mag \citep{2010AJ....140.1158T}. Finally, with the luminosities and effective temperatures obtained from the light curve solutions, the stellar radii ($R$) were computed.

To estimate the semi-major axis ($a$) of each binary system, we first combined the average fractional radii ($r_{\text{mean},1}$ and $r_{\text{mean},2}$) with their respective stellar radii ($R_1$ and $R_2$) to calculate $a_1$ and $a_2$, then took the mean of these two values. The separation between the primary and secondary stars is expressed by two values: $a_1$, measured from the primary to the secondary star, and $a_2$, measured in the opposite direction. The estimation of absolute parameters using Gaia DR3 parallax was performed separately for each star’s path, so the values of $a_1$ and $a_2$ can differ. If \(\Delta a=|a_1-a_2|\) is less than 0.1, the results are considered acceptable (\citealt{2024NewA..11002227P}). As shown in Table \ref{Tab:absolute}, the \(\Delta a\) values for the target systems are all less than 0.1. Then, using this derived semi-major axis together with the orbital period ($P$) and mass ratio ($q$), the masses of the individual stars were calculated based on Kepler's third law:
\begin{eqnarray}
M{_1}=\frac{4\pi^2a^3}{GP^2(1+q)}\label{eq:M1},\\
M{_2}=q\times{M{_1}}\label{eq:M2}.
\end{eqnarray}

The surface gravity ($g$) of each star was determined in logarithmic form, using the previously derived masses and radii. Furthermore, the orbital angular momentum ($J_0$) of the systems was computed by applying Equation \ref{eqJ0} from \citet{2006MNRAS.373.1483E}, which incorporates the total mass, mass ratio, and orbital period of the binaries.

\begin{equation}\label{eqJ0}
J_0=\frac{q}{(1+q)^2} \sqrt[3] {\frac{G^2}{2\pi}M^5P}.
\end{equation}

The absolute parameters obtained for the seven contact binary systems are listed in Table \ref{Tab:absolute}.

\begin{table*}
\renewcommand\arraystretch{1.5}
\caption{Estimated absolute parameters of the systems.}
\centering
\begin{center}
\footnotesize
\begin{tabular}{c c c c c c c c}
\hline
Parameter & J1653 & J0057 & J0159 & N3794 & V3055 Cyg & V637 Peg & Z2322\\
\hline
$M_1(M_\odot)$	&	0.29(4)	&	0.33(5)	&	0.47(3)	&	0.42(1)	&	1.09(11)	&	0.48(5)	&	0.27(4)	\\
$M_2(M_\odot)$	&	1.38(19)	&	0.99(15)	&	1.04(7)	&	0.96(1)	&	0.22(2)	&	1.20(12)	&	0.22(3)	\\
$R_1(R_\odot)$	&	0.58(5)	&	0.59(4)	&	0.74(5)	&	0.69(4)	&	1.36(9)	&	0.72(4)	&	0.56(4)	\\
$R_2(R_\odot)$	&	1.14(11)	&	0.97(7)	&	1.05(8)	&	0.99(6)	&	0.66(4)	&	1.09(6)	&	0.51(4)	\\
$L_1(L_\odot)$	&	0.19(3)	&	0.23(2)	&	0.36(4)	&	0.41(3)	&	2.34(19)	&	0.28(2)	&	0.24(3)	\\
$L_2(L_\odot)$	&	0.61(9)	&	0.51(6)	&	0.53(6)	&	0.92(8)	&	0.56(5)	&	0.53(5)	&	0.18(2)	\\
$M_{\text{bol1}}$(mag.)	&	6.55(15)	&	6.35(11)	&	5.86(12)	&	5.70(9)	&	3.82(11)	&	6.12(9)	&	6.28(12)	\\
$M_{\text{bol2}}$(mag.)	&	5.28(16)	&	5.47(12)	&	5.43(13)	&	4.84(10)	&	5.37(10)	&	5.44(10)	&	6.63(12)	\\
$log(g)_1$(cgs)	&	4.38(14)	&	4.42(12)	&	4.38(9)	&	4.38(5)	&	4.21(11)	&	4.41(9)	&	4.37(13)	\\
$log(g)_2$(cgs)	&	4.46(15)	&	4.46(12)	&	4.42(9)	&	4.43(5)	&	4.13(11)	&	4.45(9)	&	4.36(13)	\\
$a(R_\odot)$	&	2.14(11)	&	2.01(11)	&	2.21(5)	&	2.08(1)	&	2.55(9)	&	2.30(8)	&	1.39(7)	\\
$log J_0$(cgs)	&	51.44(11)	&	51.39(11)	&	51.55(5)	&	51.48(1)	&	51.30(9)	&	51.61(7)	&	50.78(12)	\\
\hline
$V_{\text{max}}$(mag.) & 13.76(11) & 13.62(9) & 13.92(9) & 13.17(8) & 13.20(9) & 12.91(8) & 12.68(8)\\
$A_V$(mag.) & 0.191(1) & 0.011(1) & 0.022(1) & 0.019(1) & 0.213(1) & 0.033(1) & 0.036(1)\\
$BC_1$(mag.)  & -0.298(19) & -0.236(13) & -0.225(13) & -0.123(10) & -0.029(6) & -0.319(13) & -0.162(10)\\
$BC_2$(mag.)  & -0.425(19) & -0.330(15) & -0.397(19) & -0.098(8) & -0.029(6) & -0.453(15) & -0.220(12)\\
$\Delta a(R_{\odot})=|a_1-a_2|$ & 0.020 & 0.010 & 0.015 & 0.006 & 0.002 & 0.007 & 0.007\\
\hline
\end{tabular}
\end{center}
\label{Tab:absolute}
\end{table*}

\vspace{0.6cm}
\section{Discussion and Conclusion}
\label{sec6}
This study presents the first detailed analysis of photometric light curves, investigate orbital period variations, and estimates the absolute parameters of eight target contact binary systems. These findings provide the foundation for the discussion and conclusions that follow:

A) Orbital period variations were analyzed for eight target systems. These changes primarily arise from factors such as apsidal motion, magnetic activity, the influence of a third body, and the transfer or loss of mass and angular momentum (\citealt{2024NewA..10502112S}).

Based on the collected and extracted times of minima, four systems exhibited linear trends (J0057, U687, V637 Peg, and Z2322), while the other four showed parabolic trends (J1653, J0159, N3794, and V3055 Cyg). We assumed that the long-term parabolic variations observed in four systems are caused by mass transfer between the two binary components. The mass transfer rate can be calculated using the following equation \citep{k1958}:
\begin{equation}
\frac{\dot{P}}{P}=-3\dot{M}(\frac{1}{M_1}-\frac{1}{M_2}).
\end{equation}

Given that the absolute parameters of these four target systems were estimated in Section 5, their mass transfer rates were subsequently calculated and are presented in Table \ref{Tab:mass-transfer}. We found that the systems exhibiting parabolic trends display a long-term decrease in their orbital periods.

B) Several studies have explored the correlation between orbital period and the temperature of the primary component ($T_1$) in contact binary systems. \cite{2011AJ....142..117S} examined this relationship using a sample of 25 such systems. Later, \cite{2017RAA....17...87Q} focused on binaries with orbital periods shorter than 0.6 days, identifying significant scatter in the data. This dispersion was attributed to the influence of tertiary companions and inaccuracies in some period measurements. \cite{2020MNRAS.493.4045J} analyzed a much larger sample and proposed two distinct linear trends, revealing a break in the $P–T_1$ diagram that divided the systems into two categories. According to their findings, early-type binaries were hotter and had shorter orbital periods, while late-type systems showed increased temperatures at longer periods. Similar patterns were reported by \cite{2020PASJ...72..103L}. \cite{2021ApJS..254...10L} also derived a linear relationship for $P–T_1$ and confirmed the presence of a break in the trend, occurring around an orbital period of 0.5 days, which is consistent with the findings of \cite{2020MNRAS.493.4045J}. \cite{2022MNRAS.510.5315P} provided additional insight by examining the relationship between orbital period ($P \leq 0.6$ days) and primary component effective temperature.

Since no temperature values were available for the Z2322 system in either the Gaia DR3 archive or the TIC database, it was necessary to estimate an initial temperature to begin the light curve analysis. To do this, we updated the empirical relationship between orbital period and the temperature of the hotter component ($T_h$) using a larger sample than those used in previous studies. The revised relation was then applied to the Z2322 system as the starting point for the modeling process.

We used a sample of 738 contact binary stars with orbital periods shorter than 0.6 days and hotter component temperatures below 8000 K. To examine and update the relationship between $P-T_h$, we applied a piecewise linear regression approach. A break point was identified in the diagram at shorter orbital periods. The location of the break point was determined by optimizing and minimizing the Residual Sum of Squares (RSS) within this framework. The resulting break point was found at an orbital period of 0.27 days. This point divided the data into two distinct segments, which may reflect underlying physical differences between contact binaries with orbital periods shorter and longer than this threshold. Confidence intervals for the linear fit parameters in each segment were estimated using ordinary least squares regression. The equations for each segment, one before the orbital period of 0.27 days (Equation \ref{eq:Segment1}) and the other between 0.27 and 0.6 days (Equation \ref{eq:Segment2}), are presented in the following:
\begin{equation}\label{eq:Segment1}
T_{h(P< 0.27)} = (12308 \pm 2016) \times P + (1919 \pm 481),
\end{equation}
\begin{equation}\label{eq:Segment2}
T_{h(0.27 < P < 0.6)} = (4594 \pm 284) \times P + (4279 \pm 109).
\end{equation}

\begin{figure*}
\centering
\includegraphics[width=0.96\textwidth]{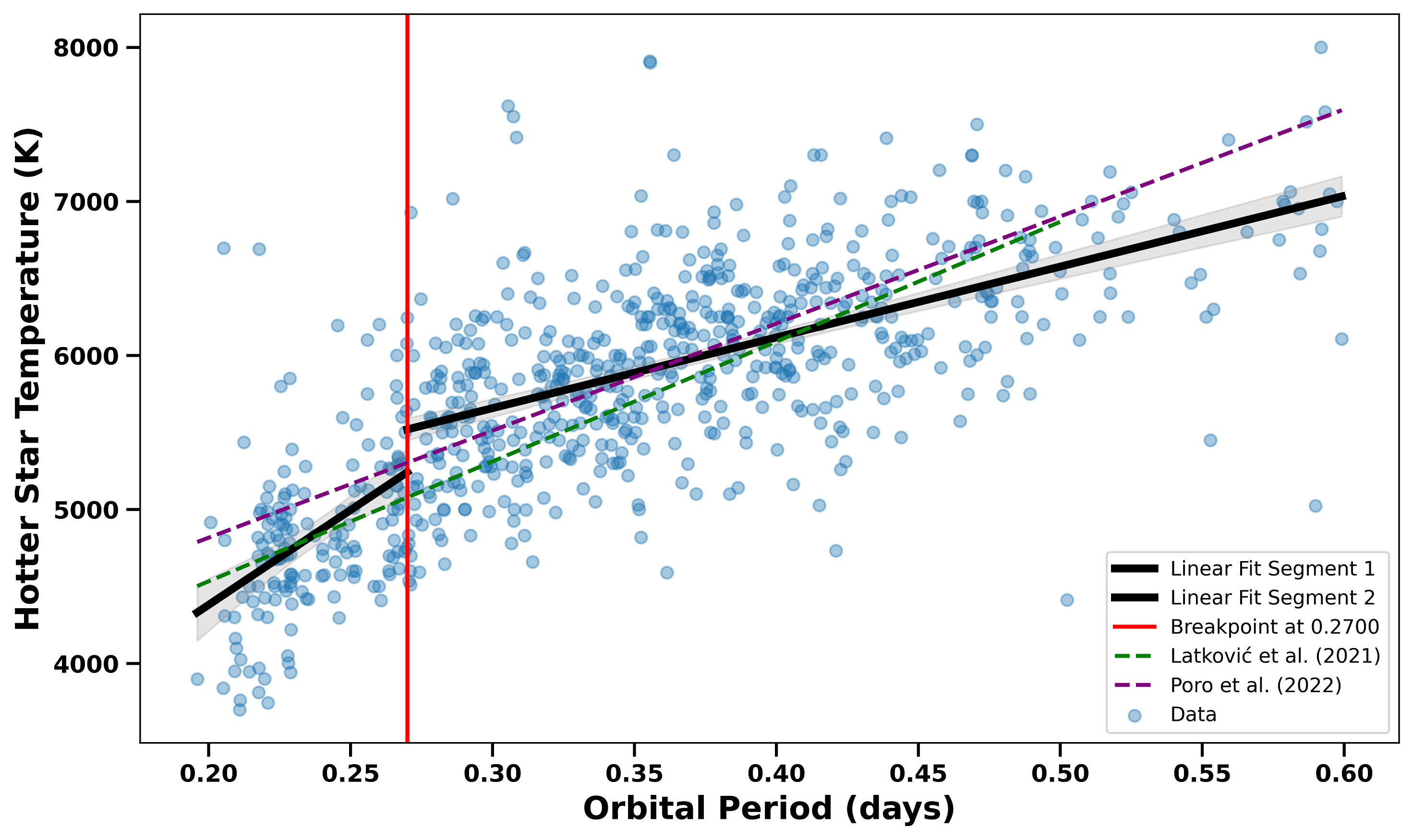}
\caption{The updated empirical relationship between orbital period and hotter component temperature in contact binary systems.}
\label{Fig:P-T}
\end{figure*}

Figure \ref{Fig:P-T} illustrates the diagram of the relationship between the orbital period and the temperature of the hotter component in the contact system, based on the sample used in this study.

C) Based on the light curve solutions, the effective temperatures of the stellar components in the target systems range from 4720 K to 6133 K. The smallest temperature difference between components was found in V3055 Cyg, with just 3 K, while U687 showed the largest difference of 411 K. Table \ref{Tab:conclusion} presents the temperature differences ($\Delta T=|T_1 - T_2|$) for all targets. The uncertainties of the temperature differences were calculated by adding the individual temperature errors in quadrature. Spectral categorizes were assigned based on temperature criteria from \cite{2000asqu.book.....C} and \cite{2018MNRAS.479.5491E}, as indicated in Table \ref{Tab:conclusion}.

Furthermore, the classification of the degree of contact was carried out using the results of the light curve solutions. The degree of contact-commonly expressed by the fillout factor ($f$)-quantifies how much the stars overfill their Roche lobes. Contact binary systems are classified according to their fillout factor as deep contact systems when \( f \geq 50\% \), medium contact systems when \( 25\% \leq f < 50\% \), and shallow contact systems when \( f < 25\% \) \citep{2022AJ....164..202L}. Therefore, three systems are classified as medium contact, while the remaining systems fall into the shallow contact category.

D) \cite{2025MNRAS.538.1427P} investigated the reliability of using Gaia DR3 parallaxes for estimating the absolute parameters of contact binary systems. They refined the acceptable difference between the primary and secondary $a$ by compiling a sample of 70 contact binaries from the literature, together with the systems analyzed in their study, for which absolute parameters had been derived from Gaia DR3 parallaxes. The reported values of $a_{1}(R_{\odot})$ and $a_{2}(R_{\odot})$ were placed on a common diagram and a statistical linear fit was performed, yielding an empirical relation. The findings of \cite{2025MNRAS.538.1427P} are in agreement with \cite{2024NewA..11002227P}. All target systems in this study have $\Delta a(R_{\odot}) < 0.1$, and therefore the results are consistent with previous studies. It is worth noting that the possible differences between $a_1$ and $a_2$ may arise from certain parameters in the light curve solution, such as $l_{1,2}$ and $r_{mean1,2}$, as well as the bolometric corrections.

E) W UMa-type contact binaries are commonly classified into two principal subtypes, designated as A-type and W-type. A defining feature of A-type systems is that the more massive star is hotter, whereas in W-type systems the more massive star is comparatively cooler \citep{1970VA.....12..217B}. Analysis by \cite{2020MNRAS.492.4112Z} revealed that these two groups follow different evolutionary pathways. This classification reflects both structural and thermal distinctions and is additionally associated with variations in angular momentum loss and the nature of mass transfer within the systems \citep{1970VA.....12..217B, 2020RAA....20..163Q}.
Based on the light curve analysis and estimated absolute parameters, four of the targets belong to the W-subtype, while three are classified as A-subtype. Due to the lack of mass estimates for the U687 system, classification of its subtype is not possible at this time. The subtype determine of seven systems are presented in Table \ref{Tab:conclusion}.

\begin{table*}
\renewcommand\arraystretch{1.2}
\caption{Some conclusions regarding the target systems include the temperature differences between the stellar components, the spectral category of each star, the fillout factor classification of each target, and the subtype classification of the systems.}
\centering
\begin{center}
\footnotesize
\begin{tabular}{c c c c c c c c c}
\hline
Parameter & J1653 & J0057 & J0159 & N3794 & U687 & V3055 Cyg & V637 Peg & Z2322\\
\hline
$\Delta T=|T_1-T_2|$ (K) & 254(69) & 224(48) & 390(52) & 116(60) & 411(43) & 3(67) & 255(38) & 183(50)\\
Sp. category & K1-K3 & K0-K2 & K0-K2 & G8-G6 & K0-K2 & F8-F8 & K2-K3 & G8-K0\\
$f$ classification & Medium & Shallow & Medium & Medium & Shallow & Shallow & Shallow & Shallow\\
Subtype & W & W & W & A & - & A & W & A\\
\hline
\end{tabular}
\end{center}
\label{Tab:conclusion}
\end{table*}

\begin{figure*}
\centering
\includegraphics[width=0.99\textwidth]{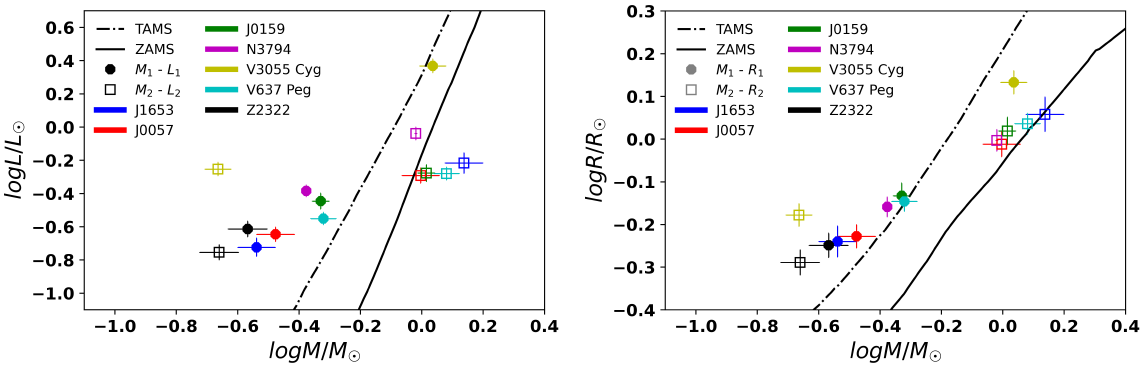}
\caption{Positions of both stellar components in each target system plotted on the $M$–$R$ and $M$–$L$ diagrams.}
\label{Fig:MLR}
\end{figure*}

F) The evolutionary stages of the targets are illustrated through logarithmic Mass-Radius ($M$-$R$) and Mass-Luminosity ($M$-$L$) diagrams based on their absolute parameters (Table \ref{Tab:absolute} and Figure \ref{Fig:MLR}). These diagrams display the components in comparison with the Zero-Age Main Sequence (ZAMS) and Terminal-Age Main Sequence (TAMS) boundaries as established by \cite{2000AAS..141..371G}.

Based on the estimated absolute parameters, the positions of the stellar companions are shown in the $M-L$ and $M-R$ diagrams (Figure \ref{Fig:MLR}). However, U687 system, for which no absolute parameter estimation was performed, is not included. Figure \ref{Fig:MLR} illustrates that the less massive components tend to be positioned near the Terminal-Age Main Sequence (TAMS), whereas the more massive stars are found closer to the Zero-Age Main Sequence (ZAMS). It is important to note, however, that contact binaries result from complex binary evolution and interactions (\citealt{2005ApJ...629.1055Y}, \citealt{2011AcA....61..139S}), which cause their evolutionary paths to differ significantly from those of isolated stars. Consequently, comparisons with single-star ZAMS and TAMS reference lines should be approached with care.

\vspace{0.6cm}
\section*{Data Availability}
Ground-based data are available in the paper's online supplement.

\vspace{0.6cm}
\section*{Acknowledgments}
This manuscript, including the observation, analysis, and writing processes, was provided by the BSN project. Ground-based observations of the target systems were conducted with the cooperation of the Observatorio Astron\'omico Nacional on the Sierra San Pedro M\'artir (OAN-SPM), Baja California, M\'exico. We used IRAF, distributed by the National Optical Observatories and operated by the Association of Universities for Research in Astronomy, Inc., under a cooperative agreement with the National Science Foundation. We used data from the European Space Agency mission Gaia\footnote{\url{http://www.cosmos.esa.int/gaia}}.

\bibliography{References}{}
\bibliographystyle{aasjournal}

\end{document}